\documentclass{PoS}
\usepackage{graphicx}

\title{On the trace anomaly in 2+1 flavor QCD }

\ShortTitle{On trace anomaly in 2+1 flavor QCD}

\author{\speaker{P. Petreczky}\thanks{for HotQCD Collaboration:
A.~Bazavov,
T.~Bhattacharya,
M.~Buchoff,
M.~Cheng,
N.~Christ,
C.~DeTar,
H.-T.~Ding,
S.~Gottlieb,
R.~Gupta,
P.~Hegde,
U.~Heller,
C.~Jung,
F.~Karsch,
E.~Laermann,
L.~Levkova,
Z.~Lin,
R.~Mawhinney,
S.~Mukherjee,
P.~Petreczky,
D.~Renfrew,
C.~Schmidt,
C.~Schroeder,
W.~Soeldner,
R.~Soltz,
R.~Sugar,
D.~Toussaint,
P.~Vranas}
\\
Physics Department, Brookhaven National Laboratory, Upton, NY 11973 USA\\
        E-mail: \email{petreczk@bnl.gov}}

\abstract{
We report on recent progress by the HotQCD collaboration in 
studying the trace anomaly at non-zero temperature in 2+1 flavor QCD on lattices with the temporal 
extent $N_\tau=4,~6$, $8$, $10$ and $12$  using the highly improved staggered 
quark~(HISQ) action as well as the asqtad action. We discuss the dependence of our lattice results on
the scale setting procedure and compare them with hadron resonance gas (HRG) calculations
as well as to resummed perturbative results.
}

\FullConference{The 30 International Symposium on Lattice Field Theory - Lattice 2012,\\
                June 24-29, 2012\\
                Cairns, Australia}

\begin{document}
\section{Introduction}

At high temperatures strongly interacting matter undergoes an transition to a new
state where quark and gluons are not subject to confinement; it shows up in a qualitative
change in the equation of state (see \cite{Petreczky:2012rq} for a review).
The most common way to calculate the QCD equation of state is
the integral method. The pressure is calculated as an integral
of the trace of the energy momentum tensor $\varepsilon-3p$ or the trace anomaly
\begin{equation}
    \frac{p}{T^4}-\frac{p_0}{T_0^4}=\int_{T_0}^T
    dT'\frac{\varepsilon - 3p}{T'^5}.
\end{equation}
The study of the trace anomaly is also interesting in its own right.
At high temperature it is determined by the running of the QCD coupling constant
$\alpha_s$, and at leading order it is proportional $\alpha_s^2 T^4$. Thus it provides a 
sensitive test of the weakly coupled nature of the
quark gluon plasma. For this reason it is also often called the
interaction measure. Thermodynamic quantities  are expected to have large cutoff effects at high
temperatures when calculated using an unimproved fermion formulation, even if the lattice spacing is small in absolute units. 
These cutoff effects arise from the distortion of the quark dispersion relation 
on the lattice and can be cured by using an improved action \cite{heller,hegde}.
On the other hand, at low and intermediate temperatures the cutoff effects in thermodynamic quantities can
be understood in terms of the cutoff effects of the hadron spectrum. In particular, the breaking of the so-called taste symmetry in the
staggered fermion formulation could be the dominant source of cutoff effects when using staggered quarks.
To control cutoff effects in pressure and other thermodynamic quantities, one first needs to understand the cutoff dependence
of the trace anomaly. In this contribution we will study in detail the cutoff
dependence of the trace anomaly and compare our lattice results with the hadron
resonance gas model at low temperature and with resummed perturbation theory
at high temperatures.
\begin{figure}[htbp]
\begin{center}
\includegraphics[width=0.495\textwidth]{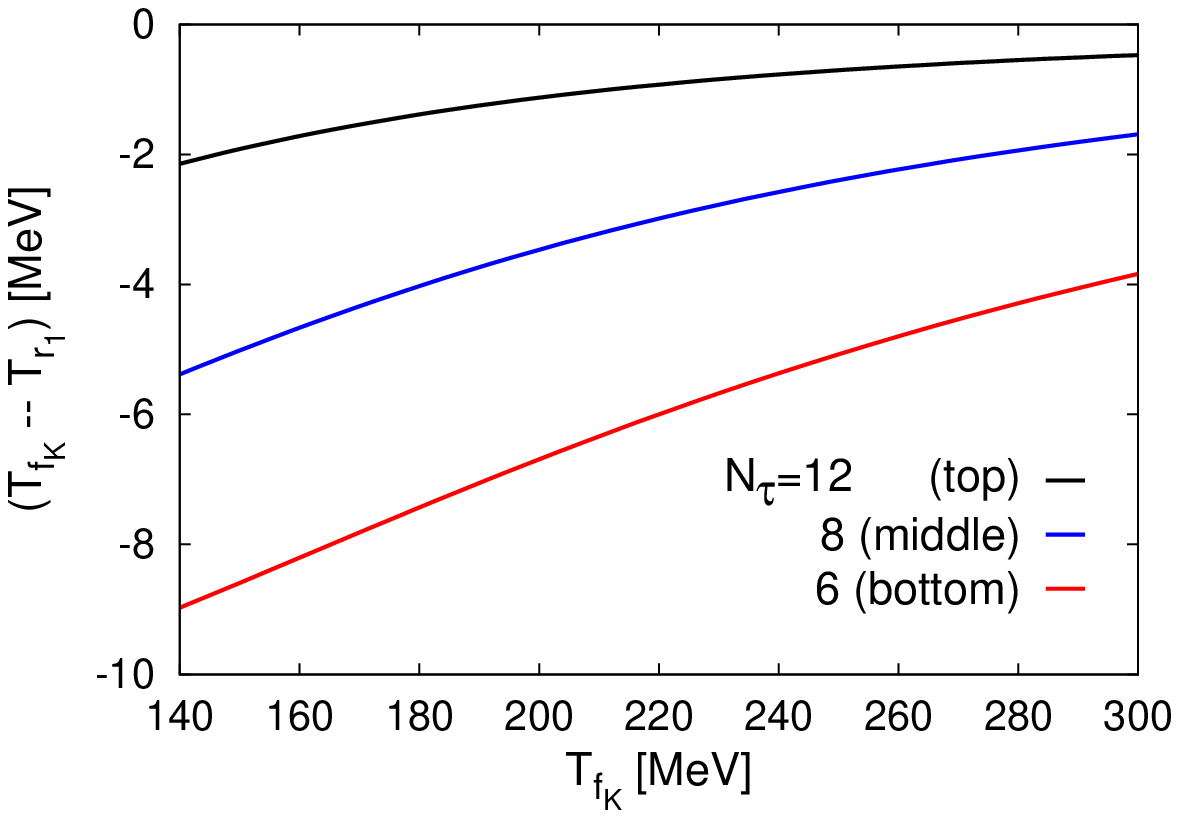}\hfill
\includegraphics[width=0.462\textwidth]{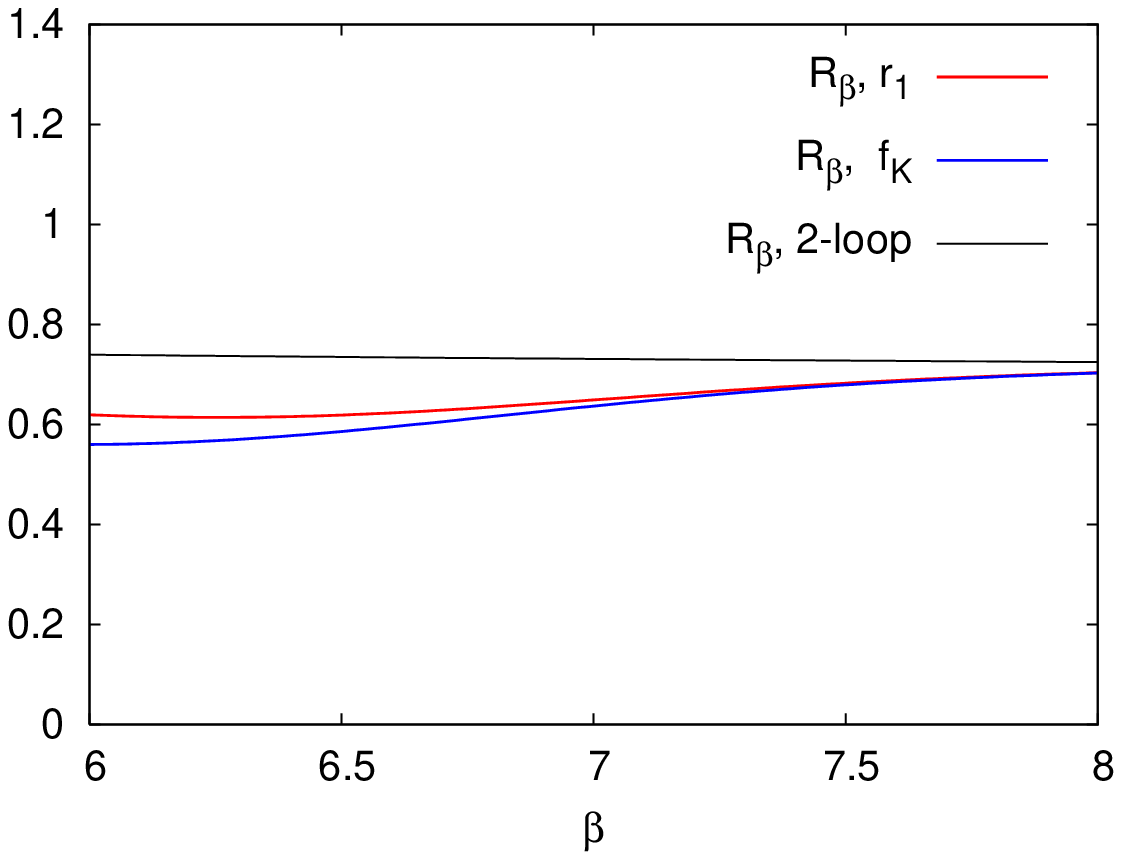}
\end{center}
\vspace{-7mm}
\caption{The difference in temperature (left) set by the $r_1$ and $f_K$
scales, described in the text. The running of the gauge coupling (right).}
\vspace{-3mm}
\label{fig_Tr1}
\end{figure}

\section{Numerical results}
We calculate the interaction measure in 2+1 flavor QCD 
using a tree-level improved action for gauge fields and the highly improved 
staggered quark (HISQ) action \cite{HPQCD}. This combination of gauge and
quark action is called the HISQ/tree action.  It can largely reduce cutoff effects
in thermodynamic quantities both at low and high temperatures \cite{proc10,tc}.

The trace anomaly can be written
in terms of the expectation values of gauge action and quark condensates as
\begin{equation}\label{tram}
  \frac{\varepsilon-3p}{T^4} = R_\beta[\langle S_{gauge}\rangle_0-
  \langle S_{gauge}\rangle_T]
  -R_\beta R_m[2m_l(\langle \bar l l\rangle_0-
  \langle \bar l l\rangle_T)+m_s(\langle \bar s s \rangle_0-
  \langle \bar s s\rangle_T)],
\end{equation}
where the subscript ``$0$'' refers to $T=0$ and ``$T$'' to finite
temperature for observables evaluated at the same value of the cutoff.
Furthermore, $R_\beta$ and  $R_m$ are the lattice beta function and mass
anomalous dimension that are defined below.
The above expression is free of ultraviolet divergences.
On a hypercubic lattice $N_s^3\times N_\tau$ the physical temperature
is set by the size of the temporal dimension and the lattice spacing
as $T=1/(N_\tau a)$. For $T=0$ calculations we use $N_\tau \geqslant N_s$,
and for $T>0$ we keep $N_s/N_\tau=4$ and at fixed $N_\tau$ vary the
lattice spacing $a$ by varying the gauge coupling $\beta=10/g^2$.
The continuum limit in this setup corresponds to $N_\tau\to\infty$.
Therefore, we carried out this study on lattices with
$N_\tau=4,~6$, $8$, $10$ and $12$. 
The strange quark mass $m_s$ is tuned
to the physical value, while the two degenerate light quarks have masses
$m_l=m_s/20$, slightly heavier than physical ($m_l\simeq m_s/27$).
In the continuum limit these light quark masses correspond to the pion mass
of about $160$ MeV.  To set the lattice spacing in physical units (fm) we use the
$r_1$ scale defined through static quark potential
\begin{equation}
r^2 \frac{d V}{d r}|_{r=r_1}=1.
\end{equation}
We use the value  $r_1=0.3106$~fm \cite{MILC_r1}.
Alternatively, the kaon decay constant, $f_K=156.1$~MeV is used to set the scale.
>From Fig.~\ref{fig_Tr1} (left) one can
see how the choice of reference observables affects the conversion
of temperature from lattice units to MeV. Over the temperature range of
interest on $N_\tau=6$ lattices the difference is within $9$~MeV, and
on $N_\tau=12$ within $2$~MeV. Another effect of using different
scales is related to the $\beta$-function and mass anomalous dimension
that enter in Eq. (\ref{tram})
\begin{equation}
R_\beta(\beta)=-a\frac{d\beta}{da},\,\,\,\,\,
R_m(\beta)=\frac{1}{m_s(\beta)}\frac{dm_s(\beta)}{d\beta},
\end{equation}
where $m_s(\beta)$ defines a line of constant physics (LCP), \textit{i.e.},
the combination of the gauge coupling and strange quark mass such that
the kaon mass (in MeV) stays approximately
constant in the whole $\beta$ range used in the simulation.
\begin{figure}[htbp]
\begin{center}
\includegraphics[width=0.495\textwidth]{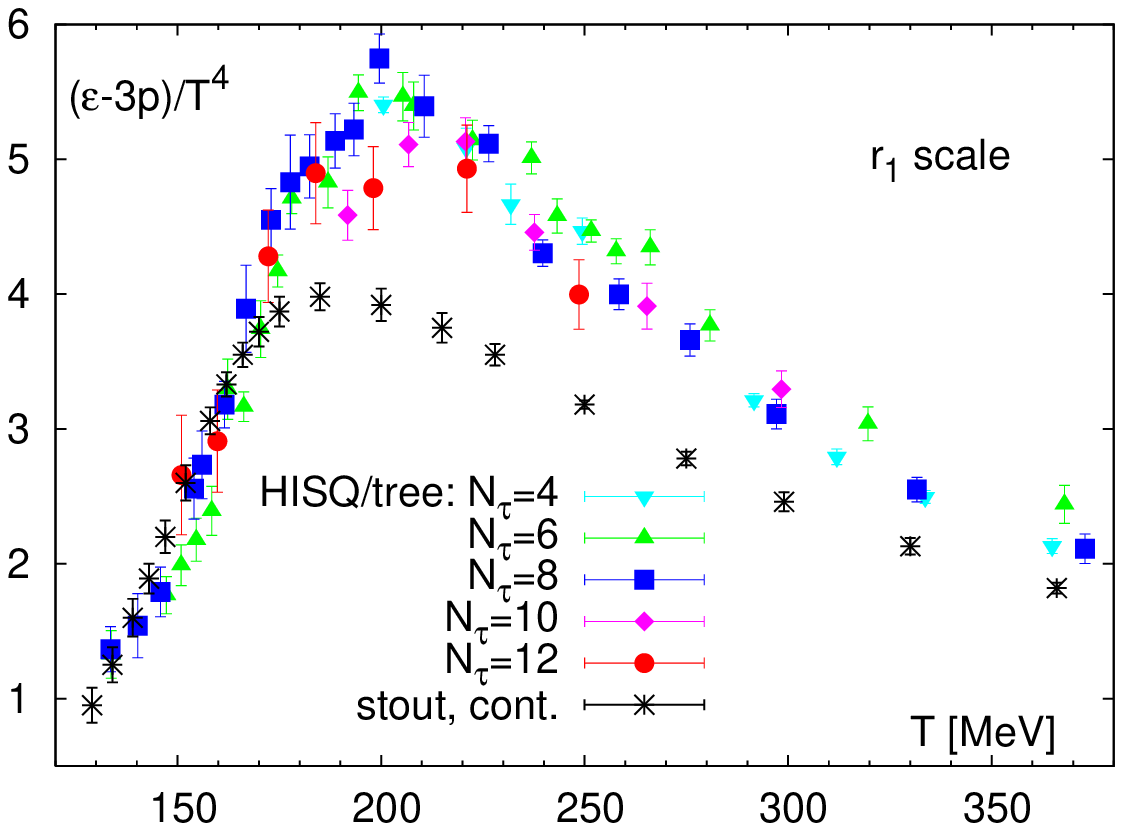}\hfill
\includegraphics[width=0.495\textwidth]{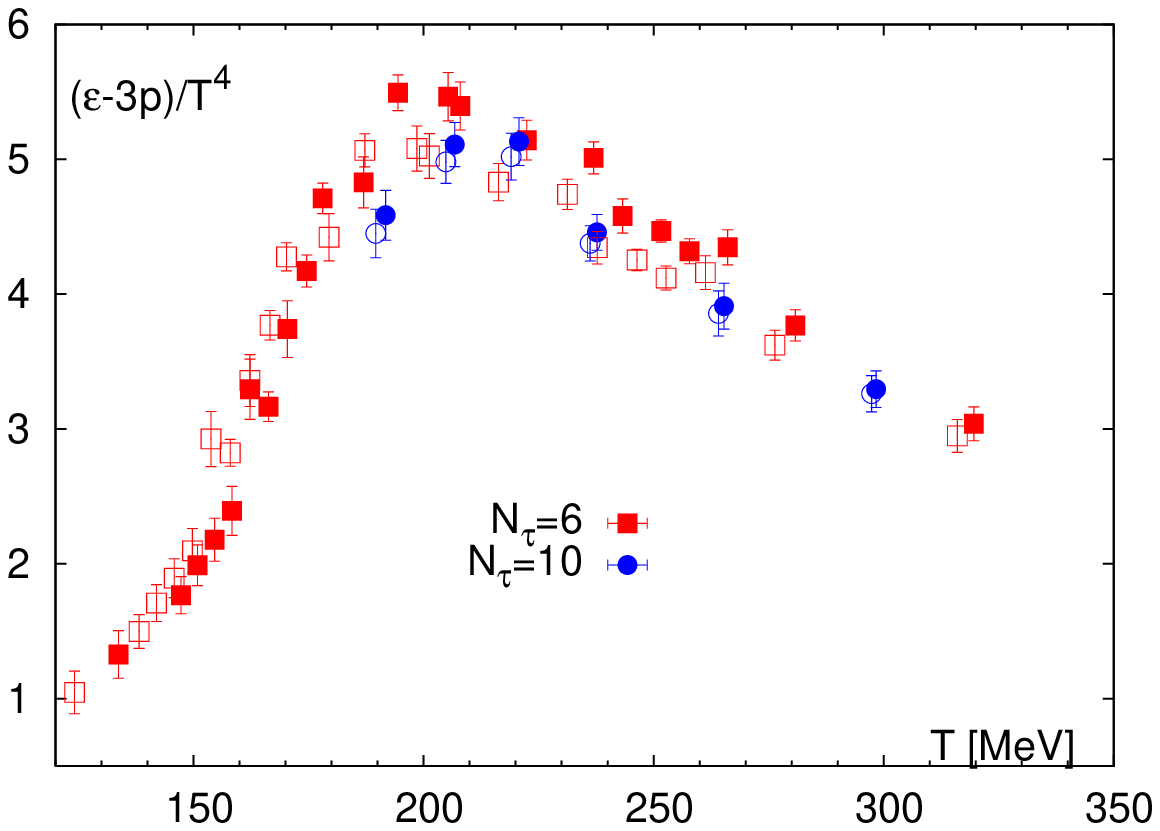}
\end{center}
\vspace{-7mm}
\caption{Comparison of the HISQ/tree interaction measure on 
$N_\tau=4,~6$, $8$, $10$ and $12$ lattices with the stout continuum
estimate~\cite{stout_eos} (left), $N_\tau=6$ and $10$ HISQ/tree
data (right). Filled (open) symbols in the right panel correspond 
to the $r_1$ ($f_K$) scale, see text. }
\vspace{-3mm}
\label{fig_EoS_hisq}
\end{figure}
$R_\beta$ for $r_1$ and $f_K$ scale is shown in Fig.~\ref{fig_Tr1} (right),
together with the 2-loop perturbative result.

The present status of the interaction measure with the HISQ/tree action is shown
in Fig.~\ref{fig_EoS_hisq} together with the stout continuum estimate of
Ref~\cite{stout_eos}. For temperature $T<180$MeV we see good agreement
between our HISQ/tree results and the stout continuum estimate. At higher
temperatures we see significant discrepancies between HISQ/tree results and
the stout results, though the discrepancies seem to disappear at temperatures
$T>350$MeV. The height of the peak decreases slightly when going from $N_{\tau}=8$
to $N_{\tau}=10$ and $12$. Thus one may expect that the discrepancies between
HISQ/tree and stout actions in the peak region will be 
reduced once the continuum extrapolation for HISQ/tree is performed.
It should be noted that in the previous calculations with p4 and asqtad
actions \cite{rbc08,hoteos,rbc10} the trace anomaly was significantly
smaller due to large cutoff effects. 

Next we study the effect of the scale setting on the trace
anomaly. In Fig.~\ref{fig_EoS_hisq} (right) the results for $N_\tau=6$ and $10$
are shown using $r_1$ and $f_K$ to set the scale. Different scale settings
affect the $N_{\tau}=6$ results, in particular in the peak region the trace
anomaly is smaller if $f_K$ is used to set the scale. However, for $N_{\tau}=10$ there
is almost no difference between the two scale setting procedures.
 
\begin{figure}[htbp]
\begin{center}
\includegraphics[width=0.495\textwidth]{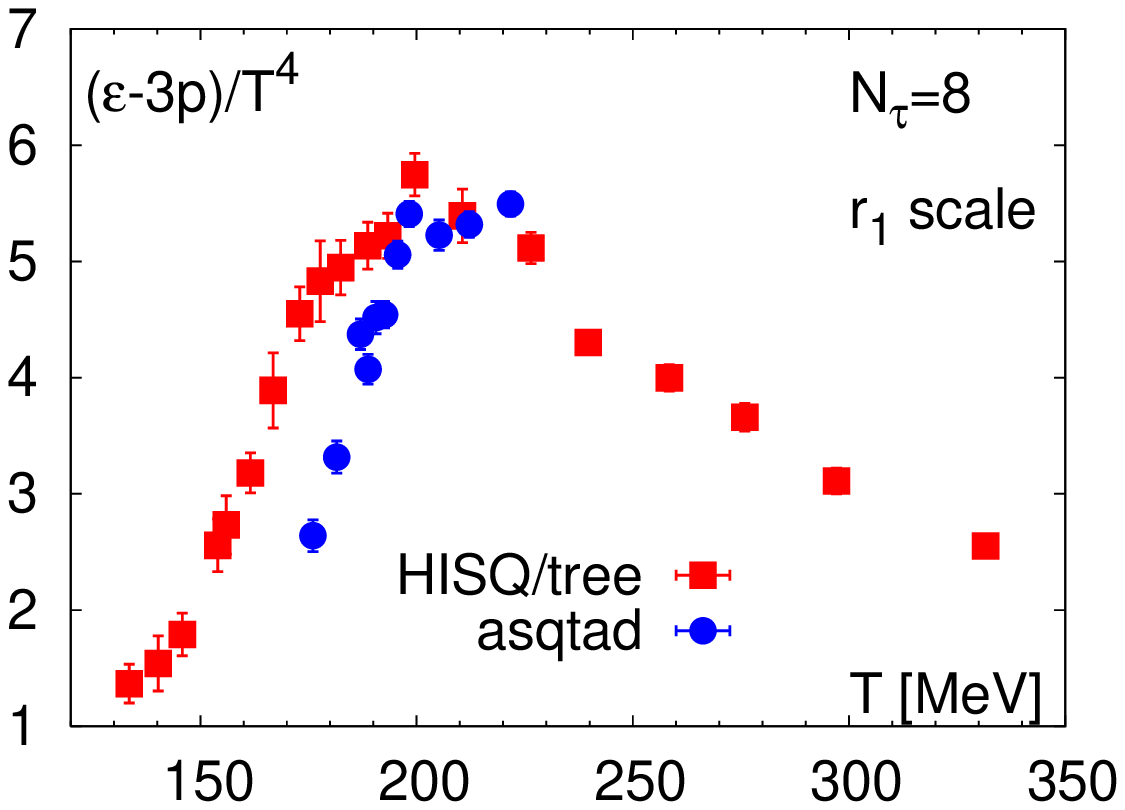}\hfill
\includegraphics[width=0.495\textwidth]{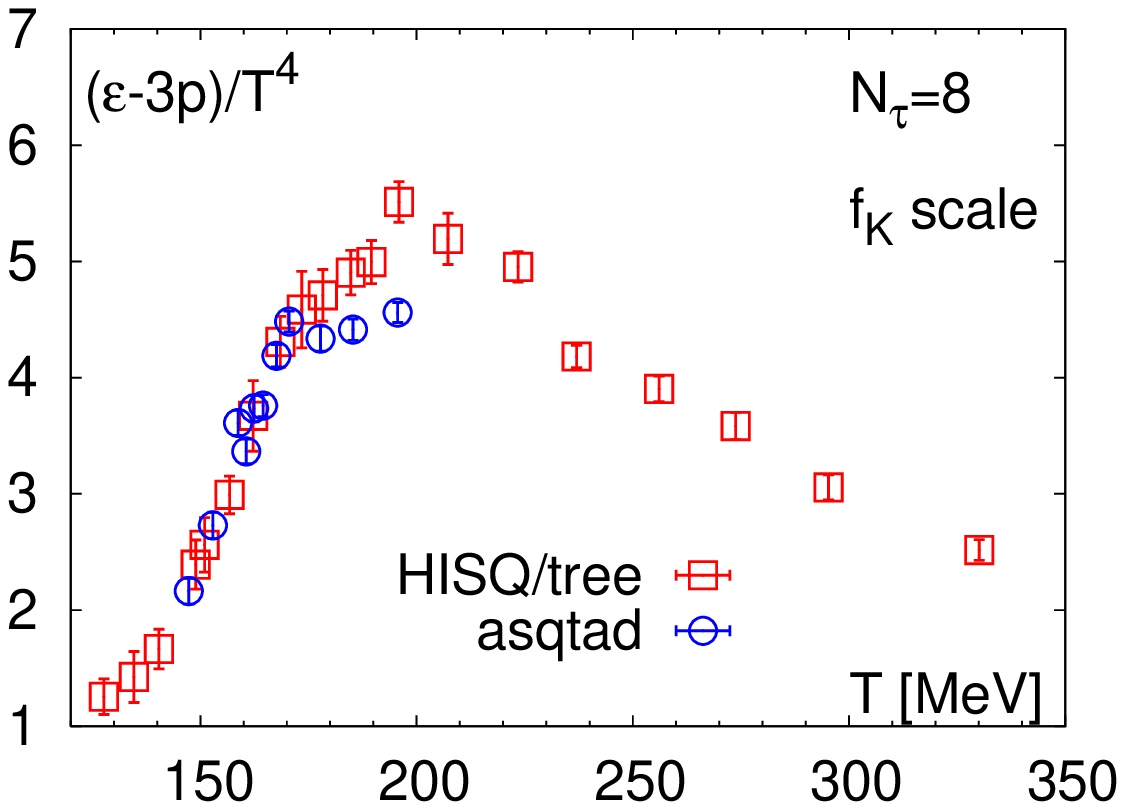}
\includegraphics[width=0.495\textwidth]{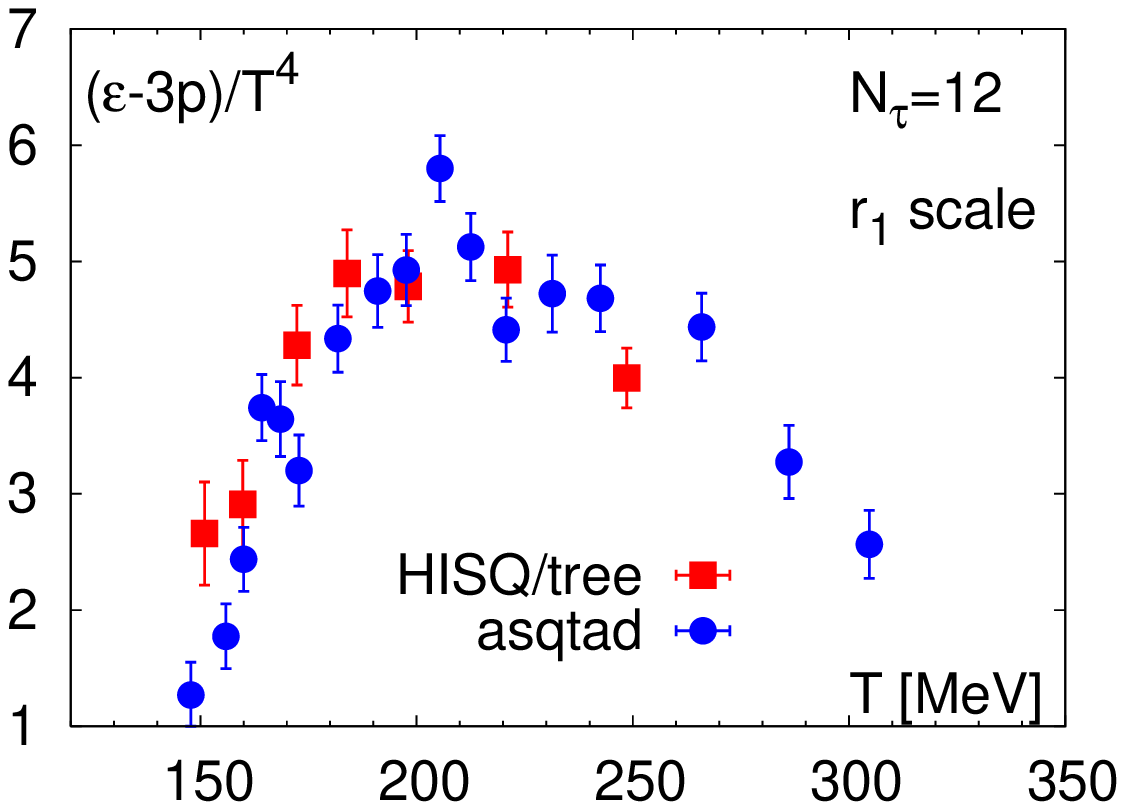}\hfill
\includegraphics[width=0.495\textwidth]{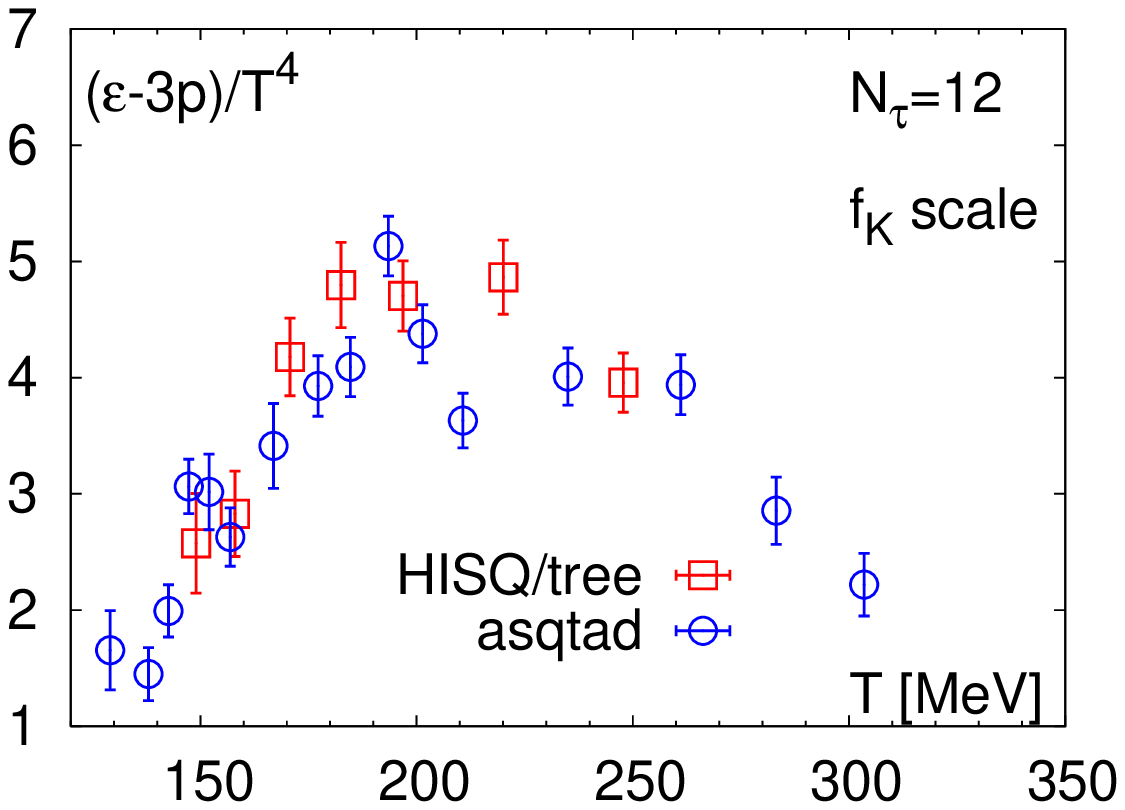}
\end{center}
\vspace{-7mm}
\caption{The interaction measure for the HISQ/tree action on
$N_\tau=8$ (top) and $N_\tau=12$ (bottom) ensembles with $r_1$ (left)
and $f_K$ (right) scale.}
\vspace{-3mm}
\label{fig_hisq_12}
\end{figure}

We also calculated the trace anomaly using the asqtad action. The expression for
the trace anomaly in terms of expectation values of local operators is given
in Ref. \cite{hoteos}. In Fig.~\ref{fig_hisq_12} we compare our results
obtained with HISQ/tree action for $N_\tau=8$ and $N_\tau=12$ using the 
scale set with both $r_1$ (filled symbols) and $f_K$ (open symbols).
If $r_1$ is used to set the scale, there is a significant discrepancy between
the HISQ/tree and the asqtad result for $T<200$ MeV for $N_{\tau}=8$. As discussed
above this is due to large cutoff effects at low temperatures when the asqtad action is used.
The discrepancy is significantly smaller for $N_{\tau}=12$, as expected. 
If we use $f_K$ to set the scale, the
discrepancies between HISQ/tree results and asqtad results are greatly reduced. Note, however,
that the peak height in the $N_{\tau}=8$ asqtad data is reduced when $f_K$ is used
to set the scale. Fig. \ref{fig_hisq_12} also shows that the difference between the two
scale setting procedures for the asqtad action is reduced in the case of $N_{\tau}=12$ data.
This is again expected. There is, however still a significant reduction in the peak
height if $f_K$ scale is used.

\begin{figure}[htbp]
\begin{center}
\includegraphics[width=0.495\textwidth]{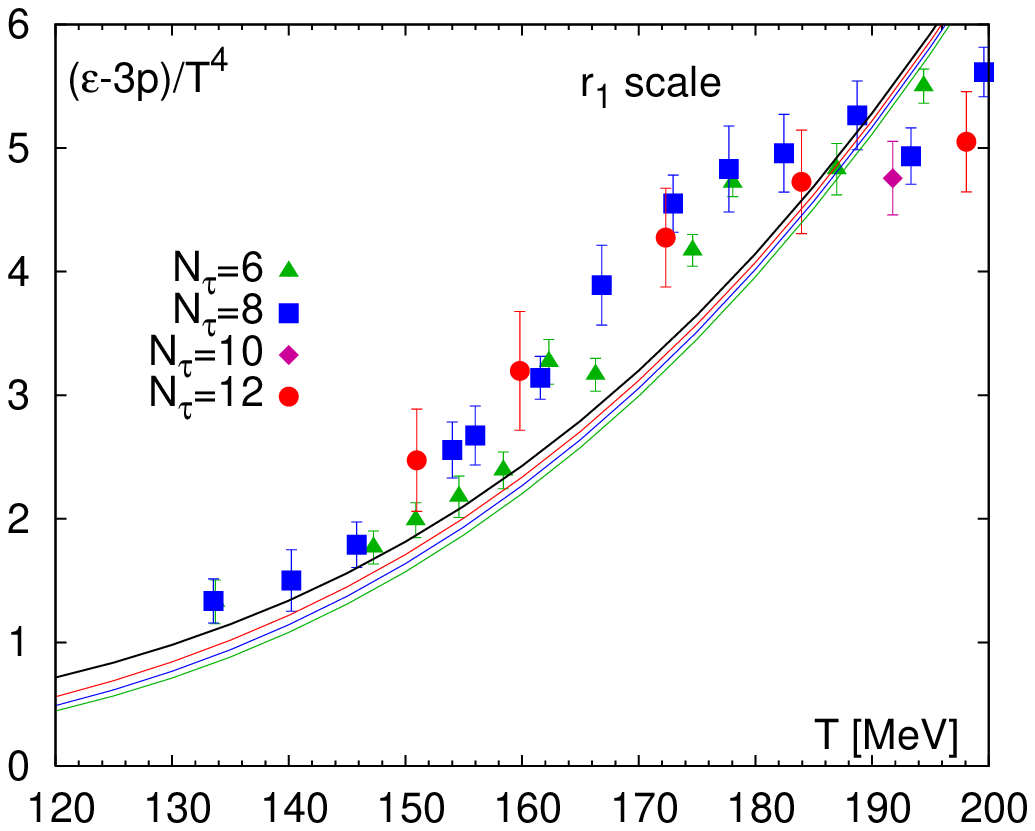}\hfill
\includegraphics[width=0.495\textwidth]{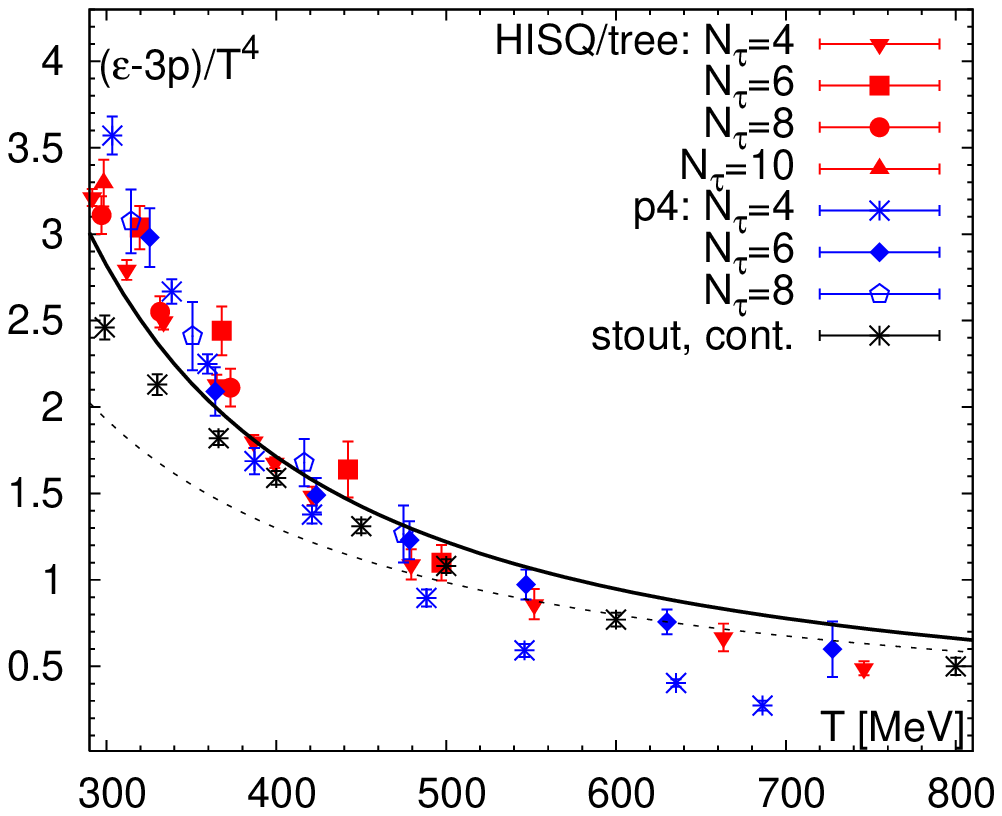}
\end{center}
\vspace{-7mm}
\caption{The interaction measure at low (left) and high (right)
temperature. The curves on the left panel represent the HRG calculation,
see text. On the right the dashed curve shows the resummed perturbative
result with 2-loop running and the solid
curve the resummed perturbative result with 1-loop running.}
\vspace{-3mm}
\label{fig_hisq_high}
\end{figure}

The low-temperature behavior of the interaction measure is shown in
Fig.~\ref{fig_hisq_high} (left) and compared to the hadron
resonance gas (HRG) calculation shown as the black line. 
At low temperatures, cutoff effects arising from the taste symmetry
breaking of the staggered fermion formulation could be significant.
To take into
account possible cutoff effects in the comparison with HRG 
we replaced the contribution of each pion and kaon with  a contribution averaged
over the sixteen different tastes of pseudo-scalar mesons. The pseudo-scalar
meson masses for each taste as function of the lattice spacing have been
previously estimated by the HotQCD collaboration \cite{tc}. In Fig.~\ref{fig_hisq_high}
we show the HRG results with the modified pseudo-scalar meson sector as colored
lines for each value of $N_{\tau}$. We use the same color coding for the lattice
results and the modified HRG result. Despite the fact that even for HISQ/tree
action the effects of taste breaking in the pseudo-scalar sector are significant,
the resulting cutoff effects are surprisingly small. They are of the same order
or smaller than the statistical errors for all $N_{\tau}$ values, including
$N_{\tau}=6$. Fig.~\ref{fig_hisq_high} shows that the lattice result
starts to disagree with the HRG model at $T\sim150-160$~MeV.

Finally, let us compare our results for the trace anomaly with earlier
calculations using the p4 action \cite{rbc08,hoteos} as well as 
with resummed perturbation theory \cite{mike}.
In Fig.~\ref{fig_hisq_high} (right) we compare our results
for the HISQ/tree action with previous results obtained with p4 action
and the stout continuum estimate. For $T>400$ MeV different lattice
data agree well with each other. The only exception is the $N_{\tau}=4$ p4 data,
which show the expected cutoff effects at high temperatures. 
The $N_{\tau}=4$ HISQ/tree data should show similar cutoff dependence, but interestingly
enough this is not observed. It is important to stress that at high temperatures the cutoff 
effects for p4 action should be small for $N_{\tau} \ge 6$. Thus the agreement between
p4 results and HISQ/tree results is expected. 
We see quite good agreement with the resummed perturbative results \cite{mike}.

\section{Conclusion}

We studied the trace anomaly in QCD using HISQ/tree and asqtad actions
and lattices with temporal extent $N_{\tau}=4,~,6,~8,~10$ and $12$.
Using different observables to set the scale, $r_1$ and $f_K$, allows
for a crude estimate of the magnitude of cutoff effects, which seem
to be small at the finest $N_\tau=12$ lattices both for HISQ/tree and
asqtad actions. Clearly to get reliable continuum results for the
trace anomaly and thus for the equation of state the calculations on
$N_\tau=10$ and $12$ lattices need to be significantly extended.
We also compared our lattice results with HRG at low temperatures
and resummed perturbative results at high temperatures. The trace
anomaly is well described by HRG for $T<150$MeV. We also find a good
agreement between lattice and resummed perturbative results at high
temperatures.

\section*{Acknowledgments}
This work has been supported by contract DE-AC02-98CH10886
with the U.S. Department of Energy.
The numerical simulations
have been performed on BlueGene/L computers at Lawrence Livermore
National Laboratory (LLNL), the New York Center for Computational
Sciences (NYCCS) at Brookhaven National Laboratory, 
US Teragrid (Texas Advanced Computing Center), 
Cray XE6 at the National Energy Research Scientific Computing Center (NERSC),
and on clusters of the USQCD collaboration in JLab and FNAL.
The calculations for asqtad actions have been performed on BG/P
computers of John von Neumann center in J\"ulich.

\end{document}